\documentclass[aps,superscriptaddress,nofootinbib]{revtex4}

\usepackage{graphicx}
\usepackage{amsfonts}
\usepackage{amssymb}
\usepackage{amsmath}
\usepackage{url} 
\newcommand{\nin}{\noindent}
\newcommand{\be}{\begin{equation}}
\newcommand{\ee}{\end{equation}}
\newcommand{\bea}{\begin{eqnarray}}
\newcommand{\eea}{\end{eqnarray}}
\newcommand{\br}{\hskip .25cm/\hskip -.25cm}

\newcommand{\nn}{\nonumber\\}

\newcommand{\ovl}{\overline}

\begin{document}

\begin{flushleft}
 KCL-PH-TH/2011-10 ~,~
CERN-PH-TH/2011-163 ~,~
LCTS/2011-10 
\end{flushleft}

\author{Jean Alexandre}
\affiliation{King's College London, Department of Physics, Strand, London WC2R 2LS, UK.}

\author{Nick E. Mavromatos}
\affiliation{King's College London, Department of Physics, Strand, London WC2R 2LS, UK.}
\affiliation{CERN, Theory Division, CH-1211  Geneva 23, Switzerland}

\title{A Lorentz-Violating Alternative to Higgs Mechanism?  }

\begin{abstract} 

We consider a four-dimensional  field-theory model with two massless fermions, coupled to an Abelian  vector field without flavour mixing, 
and to another Abelian vector field with flavour mixing. Both Abelian vectors have a Lorentz-violating kinetic term, introducing a 
Lorentz-violation mass scale $M$, from which fermions and the flavour-mixing vector get their dynamical masses, whereas the vector coupled 
without flavour mixing remains massless. When the two coupling constants have similar values in order of magnitude, a 
mass hierarchy pattern emerges, in which one fermion is very light compared to the other, whilst the vector mass is larger 
than the mass of the heavy fermion. The work presented here may be considered as a Lorentz-symmetry-Violating alternative  
to the Higgs mechanism, in the sense that no 
scalar particle (fundamental or composite) is necessary for the generation of the vector-meson mass. However,  the model  is 
not realistic given that, as a result of Lorentz Violation, the maximal (light-cone) speed seen by the fermions is smaller than 
that of the massless gauge boson (which equals the speed of light in vacuo) by an amount which is unacceptably large to be compatible 
with the current tests of Lorentz Invariance, unless the gauge couplings assume unnaturally small  values. 
Possible ways out of this phenomenological drawback are briefly discussed, postponing a detailed construction of more realistic models for future work. 

\end{abstract}

\maketitle

\section{Introduction}

We are living in the dawn of an exciting era for high energy Physics, that of LHC, where the understanding of the 
mechanism~\cite{higgs,brout,kibble} underlying the symmetry breaking sector of the Standard Model (SM) constitutes one of the 
leading objectives of the pertinent experiments. For a fundamental Higgs particle in the Lagrangian, the exclusion regions provided 
by the Tevatron~\cite{tev} improved significantly  the LEP bounds on the Standard Model Higgs mass~\cite{lep},  excluding at present 
a fundamental Higgs scalar mass $m_H$ below 115 GeV and in the region $158 < m_H < 175$~GeV~\footnote{Under the assumption of a 
fourth generation of leptons the exclusion region can be extended  to $130 < m_H < 204$~GeV~\cite{fourth}}. The LHC will soon complete 
the Higgs searches and hopefully provide a definitive(?) answer on the question regarding the existence of the elusive Higgs Boson, 
thereby  shedding  light to the Symmetry Breaking sector of the SM. The current exclusion limits from the 
LHC experiments  are (for three generation models): $m_H \in$  (155, 190)  GeV and (295, 450) GeV from the  ATLAS Collaboration~\cite{atlas}, 
and $m_H \in$  (149, 206)  GeV and (300, 440) GeV from the  CMS Collaboration~\cite{cms}. 

Although the Standard Model works very well, and the precision 
electroweak data of LEP experiments provide a very strong indication on its correctness as a low-energy physical theory describing 
electroweak and strong interactions, nevertheless its symmetry breaking sector still remains an area which may be full of surprises. 
Some of the relevant questions one may still ask is whether the Higgs Bosons is a fundamental excitation or a composite particle, 
if it exists at all. If fundamental, what is the precise form of the Higgs potential? If, composite, then what is the corresponding 
coset group structure? In this latter respect, there is a recent revival of the old technicolour-idea, which may be subjected to 
interesting phenomenological tests at LHC~\cite{grojean} in the following sense: in composite Higgs models, where the Higgs boson 
emerges as a pseudo-Goldstone boson from a strongly-interacting sector, there are additional
parameters that control the Higgs properties, which then deviate from the ones expected within the Standard Model.
Such deviations modify the LEP and Tevatron exclusion bounds and thus may significantly affect the searches for the
Higgs boson at the LHC~\footnote{Needless to say, of course, that the current exclusion bounds can be questioned already within 
the Standard Model, as a result of signal as well as theoretical uncertainties, as discussed in \cite{uncert}.}.

The assumption on the existence of a fundamental scalar particle in the SM Lagrangian, responsible for mass generation of the matter 
excitations, including the weak bosons, is one of the two major approaches to the issue. It is this approach that was put forward by 
Higgs~\cite{higgs}, and by Guralnik-Hagan and Kibble~\cite{kibble}. The other major approach to symmetry breaking, adopted by Brout and Englert~\cite{brout} is the \emph{dynamical } one, realised either by means of condensates, formed as a result of certain interactions,
as is the case of superconductivity in condensed matter systems or the chiral symmetry breaking in the four-fermion Nambu-Jona-Lasinio model~\cite{nambu}, or as a result of a \emph{selection} of a symmetry breaking \emph{solution} of a dynamical set of coupled 
field-theoretic equations (Schwinger-Dyson (SD) equations), which describe the underlying dynamics. In this approach, canonical scalar 
fields (fundamental or composite of any sort) are \emph{not} present in the Lagrangian,  and the gauge symmetry breaking is achieved 
by the dynamical appearance of a pole at zero momentum transfer $q^2 =0$ in the vacuum polarization $\Pi (q^2)$ of the gauge field, 
as a result of a massless fermion acquiring a mass through spontaneous symmetry breaking (SSB). This is the so-called Schwinger mechanism. 
In the dynamical gauge symmetry breaking approach, in contrast to the Higgs model~\cite{higgs}, there is no canonically normalised scalar 
field with an underlying  microscopic  
potential that is minimised when the field gets its vacuum expectation value (v.e.v.) that spontaneously breaks the symmetry. it goes 
without saying that the Higgs model may itself be viewed as a special case of the Schwinger mechanism in the sense that the v.e.v. of 
the Higgs scalar 
leads to tadpole contributions to the vector meson vacuum polarization that produce a pole. However, in the dynamical symmetry breaking 
scenarios, such a pole is produced by purely dynamical reasons, even in the absence of canonical scalar fields. 
The Brout and Englert approach~\cite{brout} belongs to this second class of SBB models, in the sense that, although they used in their 
analysis explicit complex scalar fields coupled to the vector meson, nevertheless they did not specify the underlying microscopic 
mechanism that yields non trivial vacuum expectation values to the gauge fields, but instead used the existence of such v.e.v. to argue 
that they provide zero-momentum-transfer poles in the vector meson polarization $\Pi(q^2=0)$, and thus a mass to the gauge boson. 

In the same class of dynamical symmetry-breaking models there belong the scalar-field-free (Higgs-less) gauge models considered in 
\cite{CornNort}, \cite{JackJohn}, which elaborated further on the ideas put forward by Brout and Englert, by avoiding any explicit use 
of fundamental scalar fields in their Lagrangians. In fact, the pertinent  models contain only fermions and gauge fields, the latter 
assumed initially massless. Then the fermions acquire mass contributions dynamically, by means of the \emph{assumption} that Nature 
\emph{selects} the appropriate symmetry breaking solution to the SD gap equations (without, however, specification of the precise 
underlying physical mechanism for such a selection, in particular there is no proof that this choice is energetically preferred in 
the sense of lowering the physical vacuum energy~\cite{CornNort}).
This choice of solutions to the SD equations results in a pole in the gauge field polarisation at zero momentum transfer, and hence 
a mass of the vector meson associated with the broken symmetry~\cite{CornNort,JackJohn}. The absence of Goldstone massless excitations 
from the physical spectrum can be demonstrated diagrammatically within the SD framework, as a result of the would-be massless poles 
in internal vector meson lines of graphs in the models of~\cite{CornNort,JackJohn} being ``\emph{eaten}'' by the longitudinal components 
of the meson fields associated with the broken symmetry group. In this way, the latter become massive by acquiring the appropriate 
degrees of freedom. 

In the original models of \cite{CornNort,JackJohn} the absence of an explicit mass scale in the Lagrangian implied that the dynamically 
generated mass depended on an arbitrary mass scale, essentially put in by hand. In subsequent works~\cite{hosek}, a mass scale $M$ 
that triggered dynamical mass generation, with a phenomenologically acceptable hierarchy between leptons and quarks, as well as weak 
vector gauge bosons, 
was provided by extra massive gauge fields, assumed to acquire a mass of order $M$.

An alternative way of generating masses for leptons and quarks has been suggested in \cite{mdyn}, based on a minimal Lorentz-Violating 
(LV) extension of Quantum Electrodynamics (QED). The model involves an explicit breaking of Lorentz symmetry, by means of higher-spatial 
derivative operators acting on the square of the Maxwell field strength, but maintains three-space rotational invariance. The higher 
derivative terms are suppressed by a mass scale $M$, which provides the scale for dynamical mass generation for the fermions. A one-loop 
analysis of the model has been performed, with the conclusion that the dynamically generated mass for the fermions was not analytic 
in the coupling constant, but instead exhibited the following dependence
\be\label{mdyn1}
m_{dyn}\simeq M\exp\left(-\frac{2\pi}{(4+\zeta)\alpha}\right)~,
\ee
where $\alpha $ is the fine structure constant, and $\zeta$ is a gauge fixing parameter. The apparent gauge fixing of (\ref{mdyn1}) 
is an artifact of the truncation used for the solution of the SD gap equation pertaining to the fermion dynamical mass. 
It has been argued in \cite{branes}, based on an extension of the so-called pinch technique argumentation~\cite{pinch} to the present LV 
model, that the true physical mass is associated with the Feynman gauge parameter $\zeta_F =0$. This argument is based on 
the cancellation of longitudinal contributions of the vector propagator, such that a
physical quantity can be obtained with the Feynman gauge, used from the beginning for the calculation of the fermion
self energy.

This presents an immediate problem: even for scales as high as the Planck mass, $M \sim 10^{19}$ GeV, the dynamically generated mass 
(\ref{mdyn1}) is unrealistically small. To obtain masses of order of the electron mass ($0.5$ MeV) one needs un-naturally high 
transplanckian mass scales $M$. This problem may be resolved by embedding the model~\cite{branes}  in microscopic brane theories, 
and applying an inverse Randall-Sundrum hierarchy. In such constructions, the seeds for Lorentz Violation (LV) may be associated with 
point-like brane defects puncturing the bulk space time in which three-dimensional brane Universes, one of which represents our world, 
propagate. The interaction of string matter with these defects induce local Lorentz violations, as a consequence of the recoil of the 
defect. The resulting low-energy effective action describing the interaction of photons with these defects contain higher spatial 
derivative terms of the LV form considered in \cite{mdyn}. The electron sector in such brane inspired models have been argued to remain 
unmodified, as a result of the fact that only electrically neutral excitations can interact non-trivially with the space-time brane defects~\cite{branes}. The advantage of embedding these models into such microscopic framework, apart from the possibility of enhancing 
the generated fermion masses to realistic values by virtue of the afore-mentioned Randall-Sundrum enhancement, also lies on the fact 
that one can obtain a physical scale $M$ which also plays the r\^ole of the UV cut-off of the low-energy theory, which is  proportional to  
\begin{equation}\label{MMs}
M \sim M_s/(g_s\, \sqrt{\sigma^2})~, 
\end{equation}
where 
$M_s$ is the string scale, $g_s$ is the string coupling assumed weak and $\sigma^2$ denotes the foam stochastic fluctuations, which is a 
free parameter in the model of \cite{branes}. 

In the string-embedded QED models~\cite{branes}, the fine structure constant (and the foam fluctuations $\sigma^2$) is also proportional 
to the $g_s^2$, and thus in the absence of string interactions $g_s \to 0$, the dynamical mass (\ref{mdyn1}) is vanishingly small. For 
non trivial string interactions, the dynamical mass remains finite, proportional to the string mass scale (\ref{MMs}). In this sense, 
the LV scenario of \cite{mdyn}) acquires a microscopic physics meaning. 
In a subsequent work~\cite{AM} to \cite{mdyn,branes}, the issue of generating dynamically fermion mass hierarchies has also been addressed.

However, in all the above scenarios, the vector boson mass was assumed not
to be generated dynamically and hence the above mechanisms fell short in providing alternative to the Higgs mechanism, not requiring 
extra fundamental scalars for gauge symmetry breaking. 
It is the point of this article to address such a question. 

The structure of the article is as follows: in the next section \ref{sec:2} we review the one-loop properties of the LV model of \cite{mdyn}, 
and concentrate on the structure of the fermion kinetic terms, including one-loop wave function renormalization. 
We demonstrate the approximate equality of the temporal and spatial components of the Dirac fermion  wave function renormalization, which 
implies 
a covariant expression for the one-loop-dressed effective kinetic fermion terms. This property supports the extension of the pinch 
technique~\cite{pinch} arguments onto this case, in order to determine the physical dynamically generated masses~\cite{branes}. Moreover, 
in this section we
also obtain a maximum speed for the fermions, which is less than that of photons, that remain massless in the model and hence they 
propagate with the speed of light \emph{in vacuo}. This is the only remnant to one-loop order of the LV nature of the model. In section 
\ref{sec:3} we proceed to 
discuss an extension of the models \cite{mdyn} to incorporate the possibility of 
dynamical mass generation of vector bosons, thereby providing us with  alternative ways to break a gauge symmetry and give mass to gauge 
bosons without the requirement of the existence of fundamental scalars.
The model consists of two massless fermions, coupled to an Abelian  vector field without flavour mixing, and to another Abelian vector 
field with flavour mixing. Both Abelian vectors have a Lorentz-violating kinetic term, introducing a Lorentz-violation mass scale $M$, 
from which fermions and the flavour-mixing vector get dynamical masses, whereas the vector coupled without flavour mixing remains massless. 
When the two coupling constants have a similar value, we find a 
fermion mass hierarchy, where one fermion is very light compared to the other. We also demonstrate dynamical mass generation for one of 
the vector bosons, whose mass is large compared to that of the heavy fermion. The other vector boson remains massless. The model may thus 
constitute a prototype where a mechanism for symmetry breaking and generation of vector meson masses, without fundamental scalars, is employed.
We also discuss the role of massless Goldstone poles in the model, associated with the dynamical breaking of the symmetry, and how 
they are absent from the physical spectrum, being ``eaten'' (as in the conventional Lorentz invariant models) by the vector meson to 
become massive. 
A discussion on Energetics of the massive ground state, as compared to the massless one,  is also discussed from the point of view of 
both the LV effective field theory and the embedding string theory model. Finally conclusions and outlook are discussed in section \ref{sec:4}.

\section{Features of a Lorentz-violating Abelian gauge theory\label{sec:2}}

We review in this section the main results of \cite{mdyn}, showing the consistency of a Lorentz-Violating (LV) field theory.\\
The Lorentz-violating Lagrangian considered in is 
\be\label{bare}
{\cal L}=-\frac{1}{4}F^{\mu\nu}\left(1-\frac{\Delta}{M^2}\right)F_{\mu\nu}+i\ovl\psi\br D\psi,
\ee
where $D_\mu=\partial_\mu+ieA_\mu$, and $\Delta=-\partial_i\partial^i=\vec\partial\cdot\vec\partial$ 
(the metric used is (1,-1,-1,-1)),
which recovers $QED$ in a covariant gauge if $M\to\infty$. 
The Lorentz-violating terms have two roles: introduce a mass scale, necessary to generate a fermion mass, 
and lead to finite gap equation, as will be seen further. 
We stress here that $M$ regularizes only loops with an internal photon line, and that
another regularization is necessary to deal with fermion loops. 
Also, the Lorentz violating modifications proposed in the Lagrangian (\ref{bare}) do not 
alter the photon dispersion relation, which remains relativistic.

We also note that more general higher order derivative extensions of $QED$ have been  
presented in \cite{kostelecky} and references therein. 
In these works, the authors consider Lorentz-violating vacuum expectation values for tensor fields,
which allow the introduction of higher order derivatives for the photon field. They explain that the Lorentz-violating Lagrangians
can be written as the Lagrangian for $QED$ in a medium, and they study for example the corresponding birefringence effects of the vacuum.
Our study in this section corresponds to a specific case of the latter models, where review~\cite{mdyn} quantum properties of a given 
Lorentz-violating 
extension of $QED$. However, the reader should also recall that, in view of the analysis of \cite{branes}, such models are embedded to 
more microscopic quantum (string) gravity constructions. 

The features reviewed here do not include the current mixing coupling of fermions which is studied in the next section, but 
these features would not essentially change, since, as we will show, the current mixing coupling involves a massive vector with mass 
$m_B$ very large compared to fermion masses. As a consequence, the corresponding excitations would imply small effects compared to those
discussed here.

\vspace{0.5cm}

\nin{\it Fermion dynamical mass}\\

Using the non-perturbative Schwinger-Dyson approach, in the ladder approximation,
we find that the fermion dynamical mass is given by~\cite{mdyn}
\be\label{mdyn}
m_{dyn}\simeq M\exp\left(-\frac{2\pi}{(4+\zeta)\alpha}\right).
\ee
Note that the expression (\ref{mdyn}) for $m_{dyn}$ is not analytic in $\alpha$, 
such that a perturbative expansion cannot lead to such a result, which
justifies the use of a non-perturbative approach.
There is an obvious dependence on the gauge parameter $\zeta$, which has a consequence on the value of $m_{dyn}$, but 
the important point is the non-analytic $\alpha$-dependence of the dynamical mass, which is not affected by the choice of gauge: 
the resulting dynamical mass is of the form $M\exp(-c/\alpha)$, where $c$ is a constant of order 1. 
This feature is known in the studies of dynamical mass generation in $QED$ in the presence of an external magnetic field \cite{mag}.
Finally, as explained in the introduction, the Feynman gauge $\zeta=0$ should be taken for the calculation of physical processes. 

\vspace{0.5cm}

\nin{\it Fermion kinetic terms}\\

The one-loop quantum corrections for the fermion kinetic term, which are different for time and
space derivatives, are
\be
i\ovl\psi \left((1+Z_0)\partial_0\gamma^0-(1+Z_1)\vec\partial\cdot\vec\gamma\right) \psi~, 
\ee
with
\bea\label{Z0Z1}
Z_0&=&-\frac{\alpha}{2\pi}\left( \ln\left( \frac{1}{\mu}\right) +4\ln2-2\right) +{\cal O}(\mu^2\ln(1/\mu)) \nn
Z_1&=&-\frac{\alpha}{2\pi}\left( \ln\left( \frac{1}{\mu}\right) +\frac{50}{9}-\frac{20}{3}\ln2\right)+{\cal O}(\mu^2\ln(1/\mu))~,
\eea
with $\mu = m_{dyn}/M$.  Note that the dominant term, proportional to $\ln(1/\mu)$, is the same for $Z_0$ and $Z_1$, since in the 
Lorentz symmetric situation ($M\to\infty$ for fixed fermion mass), we have $Z_0=Z_1$. 
Also, the coefficient $-\alpha/(2\pi)$ in front
of the dominant term $\ln(1/\mu)$ is the coefficient found in $QED$ in $4-\epsilon$ dimensions 
\be
Z_0^{QED}=Z_1^{QED}=-\frac{\alpha}{2\pi\epsilon}+\mbox{finite}~.
\ee  
An important remark should be made here: because of the result (\ref{mdyn}), the ratio $\mu$ is actually finite in the limit where
$M\to\infty$, and one could think that no counter term is necessary to absorb terms proportional to $\ln(1/\mu)$. 
But it is in fact necessary, in order 
to respect the loop structure. Indeed, if one keeps $\ln(1/\mu)$-terms in the renormalized theory, 
the would-be one-loop correction would become a tree-level one: 
\be
\alpha\ln\left( \frac{1}{\mu}\right) ={\cal O}(1)~.
\ee
Therefore, provided one treats $\ln(1/\mu)$ as a $1/\epsilon$ term in dimensional regularization, one gets the usual $QED$ one-loop 
structure. $M$ plays the role of a regulator, and the only quantity where $M$ is not set to $\infty$ is in the expression
(\ref{mdyn}) for the dynamical mass.

\vspace{0.5cm}

\nin{\it Maximum speed for fermions}\\

After redefinition of the bare parameters in the minimal substraction scheme, where
only the term proportional to $\ln(1/\mu)$ is absorbed, we find from the fermion dispersion relation that 
the product of the fermion phase velocity $v_\phi$ and group velocity $v_g$ is then
\be\label{phase}
v^2\equiv v_\phi v_g=\frac{\omega}{p}\frac{d\omega}{dp}
=1-\frac{2\alpha}{\pi}\left(\frac{34}{9}-\frac{16}{3}\ln2\right) +{\cal O}(\alpha^2)~<1~,
\ee
which shows that the effective light cone seen by fermions is consistent with causality. 
If we take $\alpha\simeq1/137$, as expected in the case of ordinary QED, we obtain for the fermion maximum speed
\be\label{vpvg}
v\simeq1-1.9\times10^{-4}~.
\ee
Unfortunately, from a phenomenological point of view, the order of magnitude of the difference (\ref{vpvg}) 
with the speed of light is unacceptably large. Indeed,
our model constitutes an explicit realization of the ideas of \cite{coleman}, where the (light-cone) maximum speed for 
fermions (electrons for concreteness in our model),  $v= c_e$,  is different, and in particular \emph{smaller} than the speed 
of photons $c_\gamma$, $c_e < c_\gamma$, which in our case, as we shall discuss below, is not modified compared to the conventional 
speed of light \emph{in vacuo}, $c$~\cite{mdyn}. Because of this, decays of ultra-high energy photons to electron-positron pairs, 
\begin{equation}\label{decays}
\gamma \to e^+ + e^-~,
\end{equation}
will be kinematically allowed for photon energies $E$ higher than 
\begin{equation}\label{threshold}
E > \frac{2 m_e c^3}{\sqrt{c_\gamma^2 - c_e^2}}~, \quad c_\gamma = c~, 
\end{equation}
If $m_e$ is replaced in the above formula by the dynamically generated physical mass (\ref{mdyn1}), 
in the Feynman gauge $\zeta = 0$~\cite{branes,pinch}, then  upon setting $c=c_\gamma$ 
and using (\ref{vpvg}), \emph{i.e}. $c_\gamma^2 - c_e^2 \sim 3.8  \times 10^{-4} \, c^2 $,  we obtain : 
\begin{equation}\label{thesh2}
E \, >  \,103 \, Mc^2\,  e^{-\pi/(2\alpha)} 
\end{equation}
If one takes for face value the dynamical mass (\ref{mdyn1}), then one may obtain  lower bounds for the cut-off $M$, by the fact that 
one observes ultra-high energy cosmic photons with energies of up to 20 TeV~\cite{coleman,sigl}. 
However, if we embed the models of \cite{mdyn} into microscopic brane models, as in \cite{branes}, in which the Randall-Sundrum 
(inverse) hierarchy 
allows for the enhancement of the dynamical mass (\ref{mdyn1}) to realistic values for the electron mass $m_e$ on our brane world, 
then  
the threshold (\ref{threshold}) and the above-mentioned fact on the observations of ultra high energy photons with energies of up 
to 20 TeV, imply
stringent constraints on the magnitude of the modification $|c_e - c_\gamma | < 10^{-15}$ ~\cite{coleman} (the absence of vacuum 
Cherenkov radiation for electrons would imply a weaker limit $c_e - c_\gamma < 5. 10^{-13}$).

Thus, phenomenologically the model, although consistent with causality, would be ruled out on the basis of the unacceptably 
large deviations (\ref{vpvg}) of the maximal speed for electrons from the speed of light \emph{in vacuo}, $c$, incompatible 
with cosmic photon (and other) observations~\cite{coleman}. However, there may be ways out which we would like to briefly 
comment upon at this stage. 

One possibility would be that the gauge groups, whose dynamics  is characterised by LV higher derivative terms, 
are beyond the Standard Model structure, as is common in string theory, for instance. In such a case their couplings would be 
free, to be fixed by phenomenology. Compatibility with the above-mentioned tests of Lorentz invariance could then be achieved, 
but unfortunately for \emph{unnaturally weak} couplings. 
Another resolution, that avoids such weak couplings,  is the extension of the model to include additional  $\gamma_5$ axial 
vector interactions among fermions  and gauge fields, as in \cite{AM}. The presence of $\gamma_5$ chirality matrix induces 
\emph{repulsive} gauge forces among the 
fermions, in contrast to the \emph{attractive } vector interactions (of QED type). Let $\alpha_{\mathcal{A}}$ be the fine 
structure constant 
of the axial interactions, and $\alpha_V$ the corresponding one for the vector interactions. The one loop corrections, then, 
that give rise to the (subdominant) finite parts of the spatial and temporal wave-function renormalization for the fermion 
(\ref{Z0Z1}), 
responsible for the maximal speed of fermions (\ref{vpvg}), may be found proportional to the combination 
$\alpha_V - \alpha_\mathcal{A}$~\cite{AM}, 
and thus could be made vanishingly small, or at least compatible with the experimental situation, upon imposing appropriate  
constraints between the axial and vector couplings.  
If such expectations are confirmed, they would imply that by enhancing appropriately the gauge structure, including axial-vector 
interactions among fermions, as is the case of the Standard Model, one may obtain a \emph{strong suppression} of the Lorentz Violating 
effects in such Lorentz-Violating extensions of the Standard Model~\cite{kostelecky} (it would be amusing to find situations implying 
the cancellation or strong suppression of the Lorentz-Violating terms, in a way resembling the gauge anomaly cancellation of the ordinary Lorentz-Invariant Standard Model). A detailed analysis along these lines is reserved for a future publication.

We next proceed to discuss the speed of photons in our model and demonstrate that it equals the speed of light \emph{in vacuo}, 
as already mentioned and used.

\vspace{0.5cm}

\nin{\it Speed of light}\\

Because $Z_0\ne Z_1$ and therefore the effective light cone seen by fermions involves the speed $v<1$, one 
might see a problem with the definition of the speed of light. We argue here that it is not the case, because of dynamical 
mass generation for the fermion. Indeed, the speed of light $c$ is defined by
\be\label{limit}
c=\lim_{m\to0}~\frac{\omega}{|\vec p|}~,
\ee
for finite momentum $\vec p$ and frequency $\omega$.
But because the fermion is always massive, $m=m_{dyn}\ne 0$, the limit (\ref{limit}) cannot be taken, and the result (\ref{vpvg}) is
not in contradiction with the speed of light.
Such a conclusion was already obtained in \cite{yukawa} for a Lifshitz-type Yukawa interaction. 
The speed of light is given by the gauge field dispersion relation, which is not modified, as we now explain.

The one-loop polarization tensor does not contain an internal photon line, such that the one-loop running coupling 
constant is the same as in $QED$. However, 
if one considers two-loop properties of the model or higher orders, the polarization tensor is affected by Lorentz violation, 
and it is necessary to make sure that the model remains consistent, especially as far as gauge invariance 
and speed of light are concerned.\\
From two loops and above, the field strength gets different corrections for time and space derivatives, and we obtain
\be
 2(1+Y_0)F_{0i}F^{0i}+~(1+Y_1)F_{ij}F^{ij}~,
\ee
where $Y_0$ and $Y_1$ represent the finite quantum corrections to the operators $F_{0i}^2$ and $F_{ij}^2$ respectively,
after absorbing the regularization terms proportional to $1/\epsilon$ or $\ln(1/\mu)$. 
In order to obtain corrections proportional to the Lorentz scalar $F_{\mu\nu}F^{\mu\nu}$, which is necessary to recover gauge 
invariance and the speed of light $c=1$, we rescale the time coordinate and the component $A_0$ as:
\be
t~\to~\frac{t}{\kappa}~~~~~\mbox{and}~~~~~~A_0~\to~\kappa A_0~~~~~\mbox{where}~~~~~~\kappa=\sqrt{\frac{1+Y_1}{1+Y_0}}~.
\ee
One can easily see then, that this rescaling is consistent with gauge invariance of the fermion sector:
\be
\ovl\psi\left( i\partial_0-eA_0\right) \gamma^0\psi~\to~\kappa~\ovl\psi\left( i\partial_0-eA_0\right) \gamma^0\psi
\ee
The factor $\kappa$, which does not appear in the space components of the covariant derivatives, will then 
contribute to the maximum speed for fermions, together with the corrections to the fermion kinetic terms, 
as explained for the one-loop case. A final identical rescaling for all the gauge field components, by the factor $\sqrt{1+Y_1}$, 
will lead to the redefinition of the coupling constant.

\section{Vector mass generation \label{sec:3}}

We consider now two vectors $A_\mu$ and $B_\mu$, with respective strengths $F_{\mu\nu}$ and $G_{\mu\nu}$.
Following the original model of \cite{CornNort}, and introducing Lorentz Violation in the vector sector 
in a way similar to \cite{mdyn}, we consider the Lagrangian
\bea\label{model}
{\cal L} = - \frac{1}{4}F_{\mu\nu}\left( 1-\frac{\Delta}{M^2}\right)F^{\mu\nu}
-\frac{1}{4}G_{\mu\nu}\left( 1-\frac{\Delta}{M^2}\right)G^{\mu\nu} +\ovl\Psi\left( i\br\partial+g_A\br A+g_B\br B\tau_2\right) \Psi~,
\eea
where $\Delta=\vec\partial\cdot\vec\partial$ is the spatial Laplacian and the fermions are flavour doublets,
\be
\Psi = \left(\begin{array}{c} \psi \\ \chi \end{array}\right)~,~~~~~~~~~~~~~
\tau_2=\left(\begin{array}{cc} 0 & -i\\i & 0\end{array}\right)~.
\ee
This theory is invariant under the gauge transformation
\bea\label{symmetries} 
&& A_\mu \rightarrow A_\mu +g_A^{-1}\partial_\mu \theta\nn
&& B_\mu \rightarrow B_\mu +g_B^{-1} \partial_\mu \theta\nn
&&\psi \rightarrow \exp(i\theta\tau_2) \psi~.
\eea
We consider the same Lorentz-violating regularization for the gauge-fixing terms,
so that the bare propagators are 
\be\label{bareD}
\left( D_{\mu\nu}^A\right)_{bare}
=\left( D_{\mu\nu}^B\right)_{bare}
=\frac{i}{1+(\vec k)^2/M^2}\left( \frac{\eta_{\mu\nu}}{k^2}+\zeta\frac{k_\mu k_\nu}{k^4}\right) ~,
\ee
where $k^\mu=(k^0,\vec k)$ and $k^2=k_0^2-(\vec k)^2$.
We shall proceed now to discuss dynamical vector boson mass generation in the model (\ref{model}) and the 
associated patterns of the breaking of the gauge symmetry (\ref{symmetries}).

\subsection{Dressed fermion sector}

In this subsection we shall demonstrate that the fermions in the model (\ref{model})  acquire a dynamical mass, with the mechanism 
already described in \cite{mdyn} for the case of a single fermion interacting with a single gauge field through a vector coupling.
Neglecting the wave function renormalization of the form (\ref{Z0Z1}), we look for a fermion self-energy of the form
\be
\Sigma=m_1+m_2\tau_3~~~~\mbox{with}~~~~\tau_3=\left(\begin{array}{cc} 1 & 0\\0 & -1\end{array}\right)~,
\ee
where $m_1$ and $m_2$ are generated dynamically and are related to the fermion dynamical masses $m_\psi$ and $m_\chi$ by
\be\label{mamb}
m_\psi=m_1+m_2~~~~\mbox{and}~~~~m_\chi=m_1-m_2~.
\ee 
The dressed fermion propagator is therefore
\bea
G&=&\frac{i}{\br p+m_1+m_2\tau_3} =i\frac{\br p+m_1-m_2\tau_3}{(\br +m_1)^2-m_2^2}\nonumber \\
&=&i\frac{(m_1-m_2\tau_3)(-p^2-m_1^2+m_2^2)+2m_1p^2}{4m_1^2p^2-(p^2+m_1^2-m_2^2)^2} 
+ i\br p~\frac{-p^2+m_1^2+m_2^2-2m_1m_2\tau_3}{4m_1^2p^2-(p^2+m_1^2-m_2^2)^2} ~,
\eea
where $p^\mu=(p^0,\vec p)$ and $p^2=p_0^2-(\vec p)^2$.
We next proceed to discuss the dynamical generation of mass for the vector Boson $B_\mu$ in the model, which constitutes 
the main topic of our discussion in this work.

\subsection{Dressed B-vector sector}

As explained in \cite{CornNort}, the vector $B_\mu$ will dynamically acquire the mass $m_B$, and we review here the 
non-perturbative mechanism, in the Lorentz-symmetric case.\\
The Ward identity corresponding to the interaction with the vector $B_\mu$ is 
\be\label{WardB}
g_B^{-1}k^\mu\Gamma^B_\mu(p-k,p)=\tau_2G^{-1}(p)-G^{-1}(p-k)\tau_2~,
\ee
where $k_\mu$ is the momentum of the vector meson\footnote{Note that this ward identity is the same for the LV model (\ref{model}),
since its derivation is independent of the form of the bare vector propagator. The difference with the Lorentz symmetric case
is that the dressed $n$-point functions depend on $p^0$ and $\vec p$ independently, and not only through the combination 
$p_0^2-(\vec p)^2$.}. As a consequence, if the dynamical mass $m_2$ is indeed generated, we obtain
\be\label{anticom}
\left[ \tau_2,G^{-1}(p)\right] \ne0~,
\ee 
and the limit $k_\mu\to0$ gives a non-vanishing right hand side of the Ward identity (\ref{WardB}), which is possible only if 
the dressed vertex has a pole at $k_\mu=0$. One can then decompose the vertex into a regular part $\Gamma_\mu^{reg}$ and a singular 
part $\Gamma^{sing}$ to write
\be\label{regsing}
\Gamma^B_\mu(p,k)=\Gamma_\mu^{reg}(p,k)+\tau_1\frac{k_\mu}{k^2}\Gamma^{sing}(p,k)~,~~~~~
\mbox{where}~~~~~\tau_1=\left(\begin{array}{cc} 0 & 1\\1 & 0\end{array}\right)~.
\ee
The reader should notice here that Lorentz invariance has been assumed for this decomposition, since the singular part is 
assumed proportional to $k^\mu$. We shall come back to this important point in the next subsection, when we discuss the 
cancellation of massless poles from scattering amplitude, as required by the Goldstone theorem. 

For the moment, based on the above considerations, we consider the Schwinger-Dyson equation for the vector propagator
\be\label{SDB}
\Pi_{\mu\nu}=g_B^2\mbox{Tr}\int_p\Gamma_\mu^B(p-k,p)G(p)\gamma_\nu\tau_2G(p-k)~,
\ee
from which one finds the transversality of the polarization tensor $k^\mu\Pi_{\mu\nu}=0$. As a consequence, the
$B_\mu$-vector propagator is of the form
\be\label{DB1}
D^B_{\mu\nu}(k)=\frac{1}{1 + (\vec k)^2/M^2}\, \left(\frac{1}{k^2-k^2\Pi_B(k)}\left( \eta_{\mu\nu}-\frac{k_\mu k_\nu}{k^2}\right) 
+(\zeta+1)\frac{k_\mu k_\nu}{k^4}\right)~.
\ee
From this last expression, we see that a vector mass $m_B$ can be generated if $\Pi_B(k)$ has a pole for $k^\mu=0$.\\
The next step is to differentiate the Ward identity (\ref{WardB}) with respect to $k_\mu$, and plug the decomposition 
(\ref{regsing}) into the Schwinger-Dyson equation (\ref{SDB}) to obtain
\be\label{integral} 
\lim_{k\to0}\left\lbrace k^2\Pi_B(k)\right\rbrace 
=\frac{g_B^2}{4}\mbox{Tr}\int_p\gamma^\mu\Big( \frac{\partial}{\partial p^\mu} G(p)+G(p)\Gamma_\mu^{reg}(p,p)G(p)\tau_2\Big)
=m_B^2~,
\ee
where one notices that the singular part $\Gamma^{sing}$ does not appear. 
We note that the pole structure retains its Lorentz-Invariant form in our minimal LV theory, 
since the Lorentz violation in the vector meson propagator enters only via non-singular factors $1/(1 + (\vec k)^2/M^2)$. 
As we shall discuss in the next section, the $1/k^2$ pole is due to the exchange of a massless spin-zero Goldstone particle, 
associated with the breaking of the $B$-vector local rotation (gauge) symmetry generated by $\tau_2$ (\ref{symmetries}). 

We next remark that the integral (\ref{integral}) is  
regularized in \cite{CornNort} with an additional power of the momentum in the fermion propagator:
\be
G^{-1}(p)=\br p+m_1+m_2\tau_3\left(\frac{m_2}{p^2} \right)^\eta~, 
\ee
where the limit $\eta\to0$ will be taken after integration over momentum. This limit is taken in such a way that the ratio
\be\label{limitepsilon}
\frac{\eta}{g_A-g_B}\to ~\mbox{finite}~,
\ee
and the resulting mass for the vector $B_\mu$ is
\be\label{mB}
m_B^2=\frac{2}{3}\frac{g_B^2}{g_A^2-g_B^2}(\Delta m)^2=\frac{8}{3}\frac{g_B^2}{g_A^2-g_B^2}m_2^2~,
\ee
where $\Delta m=2m_2$ is the fermion mass splitting.\\
Although we have a Lorentz-violating 
theory, we will assume that the vector mass $m_B$ is well approximated by the expression (\ref{mB}).
Also, we will neglect corrections to the Lorentz-violating factor, so that the dressed propagator we will use 
for the vector $B_\mu$ has the form:
\be\label{DB}
D^B_{\mu\nu}=\frac{1}{1+(\vec k)^2/M^2}\left( \frac{\eta_{\mu\nu}-k_\mu k_\nu/k^2}{k^2-m_B^2}+(\zeta+1)\frac{k_\mu k_\nu}{k^4}\right) ~.
\ee
We shall use this propagator to discuss fermion mass generation as a consistent solution to the pertinent Schwinger-Dyson equations 
and then demonstrate the associated mass hierarchy. Before doing so, though, we consider it as essential to discuss the cancellation 
of massless Goldstone poles from physical scattering amplitudes of this model, which is an essential feature of all such dynamical 
gauge symmetry breaking models~\cite{JackJohn,CornNort}.

\subsection{Absence of Goldstone Poles from the Physical Spectrum}

The dynamical breaking of the symmetry associated with the masslessness of the vector boson, $B_\mu$, would imply, according to 
Goldstone's theorem, a massless pole. However, such a massless pole is not present in the physical spectrum as it is ``eaten'' by 
the corresponding vector boson to become massive. This is demonstrated explicitly in the Higgless models dynamically in \cite{JackJohn},
\cite{CornNort} and the proof can be extended in our minimal LV models, by making use of the form of the decomposition (\ref{regsing}) 
of the vertex function into a regular and a pole part. The Lorentz invariant form of the dressed vertex, assumed in our minimal LV model, 
is crucial for the demonstration of the absence of massless Goldstone poles  from the physical scattering amplitudes. 
 
\begin{figure}[ht]
\centering
\includegraphics[width=9cm]{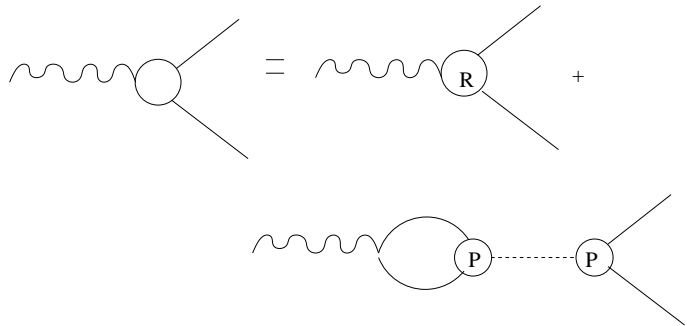} 
\caption{The decomposition (\ref{regsing}) of the the $B$-meson-fermion vertex function into a regular and a pole (singular) part; 
the pole is attributed to the exchange of a massless Goldtone spin-zero excitation which though is absent from the physical spectrum 
(scattering amplitude) due to delicate cancellations between singular vertex and propagation poles. The notation ``R'' stands for 
Regular (non-singular) parts and ``P'' for pole (singular) graphs.  }%
\label{fig:polevertex}
\end{figure}

We follows here the argument given in \cite{JackJohn} for the graph representing fermion scattering, and we concentrate for simplicity 
on the graphs involving the $B_\mu$ vector meson only, which are relevant for the symmetry breaking patterns. The presence of the 
(massless) vector meson $A_\mu$ does not affect our arguments.  The  graph representing the vertex decomposition (\ref{regsing}) is 
given in fig.~\ref{fig:polevertex}.  The singular parts of the vertex function is due to the exchange of the massless spin-zero scalar) 
Goldstone modes, which exist only in internal lines and they do not couple directly to the $B$-vector mesons.

\begin{figure}[ht]
\centering
\includegraphics[width=9cm]{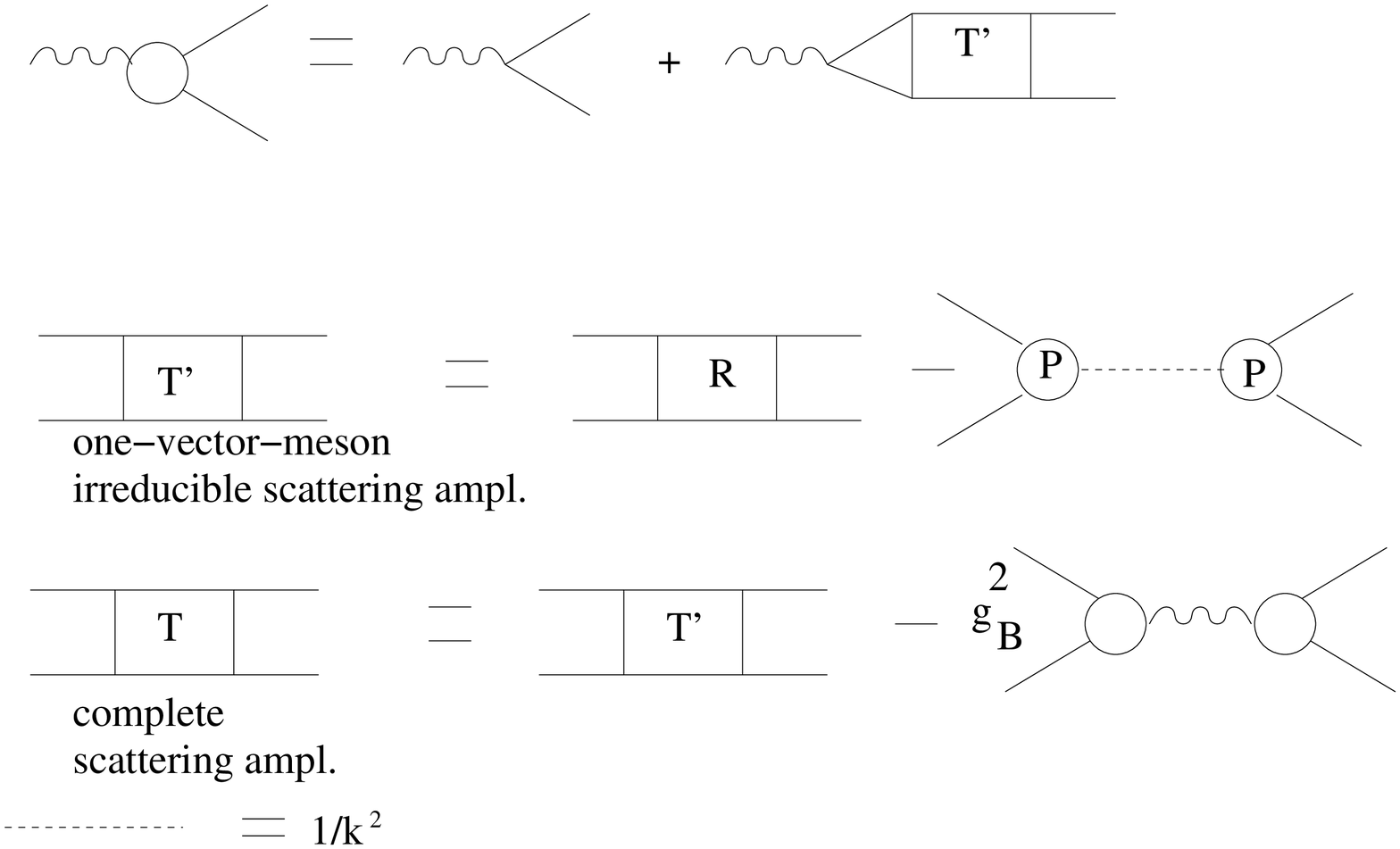} 
\caption{The upper figure shows the quantum corrections to the $B$-meson fermion vertex function, 
including the one-vector-meson-irreducible parts $T'$. The middle figure shows the decomposition of $T'$ into a regular (``R'') and 
a pole (``P'') part, due to the Goldstone scalar exchange (denoted by a dash line, corresponding to the $1/k^2$ pole structure). 
The lower figure gives the total scattering amplitude $T$  of two fermions in terms of the $T'$ part and the one $B$-vector-meson  exchange.}
\label{fig:scatampl}
\end{figure}

The total scattering amplitude $T$ contains a one-vector-meson-irreducible part ($T'$) and a one-vector-meson exchange part, and is 
depicted in fig.~\ref{fig:scatampl}. The external fermion lines are\emph{ on-shell}. Notice that in our minimal LV scenario the LV 
terms appear as non singular factors 
$1/(1 + (\vec k)^2/M^2)$ in the vector meson propagators (\ref{DB1}), but Lorentz invariant structure is assumed for the vertex functions, 
as a consequence of the form of the Ward identity (\ref{WardB}), as discussed above and in \cite{branes}.  In fact it is this assumption 
that allows us to extend the pinch technique arguments of ordinary Lorentz invariant theories~\cite{pinch}, to our minimal LV scenario, 
in order to argue that physical gauge invariant results are obtained by ignoring the longitudinal parts of the vector meson propagators. 

The $T'$ part contains a pole due to the Goldstone massless scalar exchange (denoted by a dash line in the graphs). Such exchanges are 
responsible for the $1/k^2$ poles structure, where $k$ is the intermediate momentum transfer. These diagrammatic structures follow 
directly from the pertinent Schwinger-Dyson equations, upon using the decomposition (\ref{regsing}). 

\begin{figure}[ht]
\centering
\includegraphics[width=9cm]{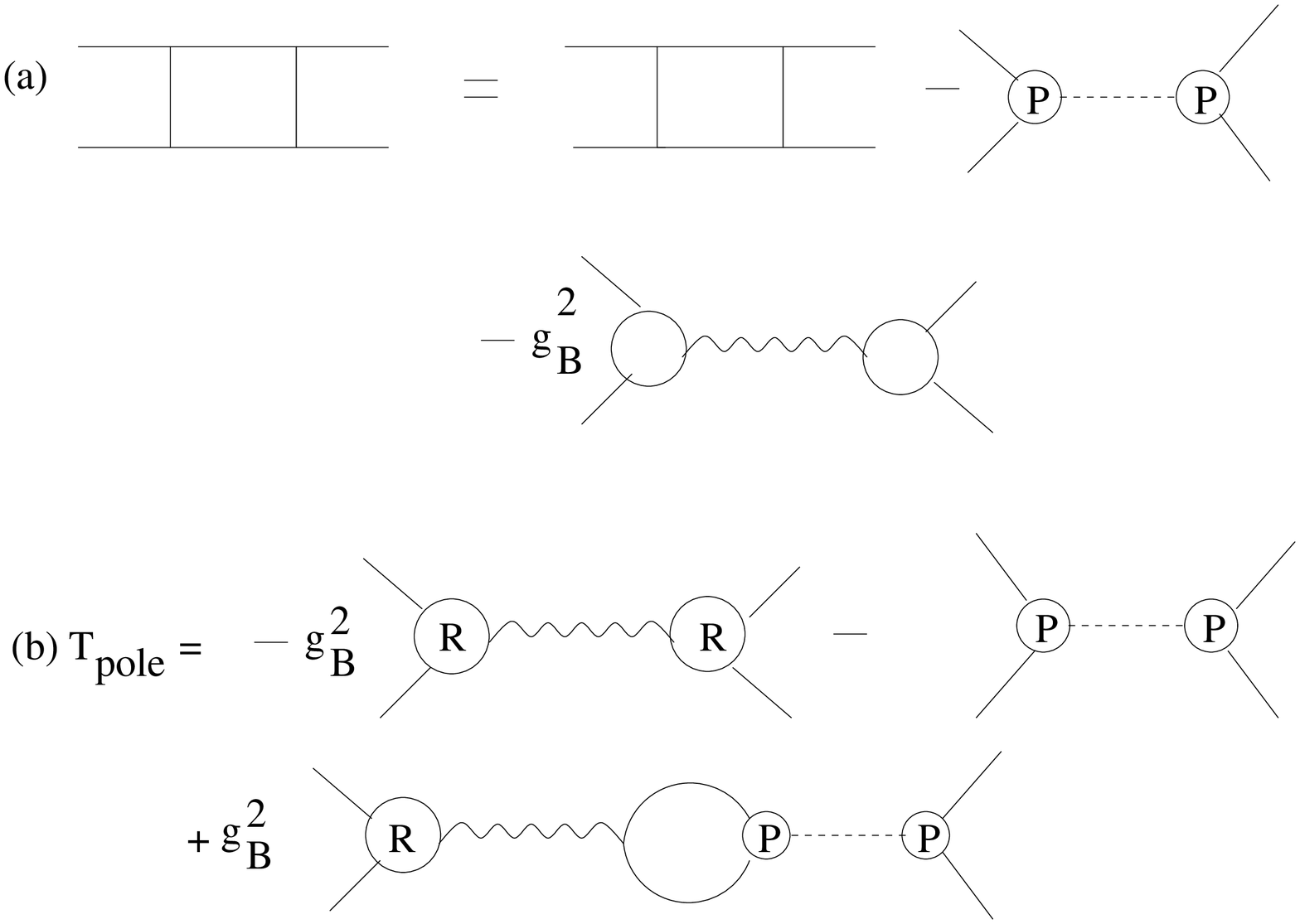} 
\caption{The upper figure (a) shows the complete scattering amplitude for two fermion scattering $T$, while the lower figure (b) 
details its pole structure, using the vertex decomposition (\ref{regsing}), \emph{cf}. fig.~\ref{fig:polevertex}.}%
\label{fig:completescat}
\end{figure}

The complete physical scattering amplitude $T$ is represented in the upper graph (a) of fig.~\ref{fig:scatampl}, and its pole structure 
is detailed in the lower figure (b). In the last  graph of figure (b) we may 
use again the decomposition of the vertex function (\ref{regsing}) in order to express the regular part of the vertex that appears on the 
left hand side of the vector meson in terms of the full vertex and the singular part: 
$\Gamma_\mu^{reg}  = \Gamma_\mu^B - \tau_1 \frac{k_\mu}{k^2} \Gamma^{sing}$.
Since the external fermions are on-shell, the full vertex is transverse, $k^\mu \Gamma^B_\mu =0 $, such that
only the contribution proportional to $\eta_{\mu\nu}$ of the vector propagator plays a role in the total amplitude $T$,
and not the contribution proportional to $k^\mu k^\nu$. We write this contribution $\eta_{\mu\nu}D_B$. The 
$B$-vector exchange is proportional to $\Gamma^\mu_B\eta_{\mu\nu}D_B\Gamma^\nu_B$, with the following pole structure
\be\label{pole1}
\sim~~\frac{k^\mu}{k^2} \Gamma^{sing}(p',k)~\frac{1}{1+(\vec k)^2/M^2}~\frac{i\eta_{\mu\nu}}{k^2-k^2\Pi_B(k)}
~\frac{k^\nu}{k^2} \Gamma^{sing}(p,k)
~~~\underrightarrow{k\to0}~~~
-\frac{i}{m_B^2}\Gamma^{sing}(p,0)\frac{1}{k^2}\Gamma^{sing}(p,0)~.
\ee
The latter pole structure is cancelled by the pole corresponding to the exchange of the scalar excitation, 
with propagator $i/k^2$ and which has an effective derivative coupling $\xi k^\mu$ to the fermions \cite{JackJohn}, with
\be
\xi^2=\lim_{k\to0}\left\lbrace \frac{1}{k^2\Pi_B(k)}\right\rbrace =\frac{1}{m_B^2}~.
\ee
Indeed, this graph leads to the following pole structure
\be
\sim~~\xi^2k_\mu k_\nu~\frac{k^\mu}{k^2}\Gamma^{sing}(p',k)~\frac{i}{k^2}~\frac{k^\nu}{k^2}\Gamma^{sing}(p,k)
~~~\underrightarrow{k\to0}~~~
\frac{1}{m_B^2}\Gamma^{sing}(p,0)~\frac{i}{k^2}~\Gamma^{sing}(p,0)~,
\ee
which is the opposite of the contribution (\ref{pole1}). As a consequence, the effective scalar excitation
does not appear in the spectrum, and decouples from the system. This cancellation is depicted in fig.~\ref{fig:completescat}.\\
The same proof can be extended to all physical processes. 
The extension  of the pinch technique arguments of the Lorentz Invariant case to this minimal LV model~\cite{pinch,branes} ensures the 
contribution only of the transverse part of the propagator, as used above.  Hence the spin-zero Goldstone massless excitations disappear 
from the physical spectrum of the model, in agreement with the general arguments on dynamical breaking of the symmetry. 

Finally, we remark that here the would-be Goldstone mode refers to the spontaneous breaking of the gauge symmetry (\ref{symmetries}). 
The Lorentz Invariance is broken explicitly in the model, due to the presence of the higher-derivative terms (in terms of the microscopic 
models of ref.~\cite{branes}, the explicit breaking of Lorentz symmetry is provided by the recoiling point-like D$0$-brane (D-particle) 
defects in space-time).

\subsection{Fermion gap equations}

The Schwinger-Dyson equation for the fermion propagator is
\be\label{SD}
\Sigma=ig_A\int\gamma^\mu G\Gamma_A^\nu D_{\mu\nu}^A
+ig_B\int\gamma^\mu\tau_2 G\Gamma_B^\nu D_{\mu\nu}^B~,
\ee
where $D_{\mu\nu}^A$ and $D_{\mu\nu}^B$ are the dressed vector propagators, $\Gamma_A^\nu$ and $\Gamma_B^\nu$ are the corresponding
dressed vertices, and $G$ is the dressed fermion propagator. We will consider the so-called ladder approximation, where the dressed
vertices are assumed to be the same as the bare ones:
\be
\Gamma_A^\nu\simeq ig_A\gamma^\nu~~~~\mbox{and}~~~~\Gamma_B^\nu\simeq ig_B\gamma^\nu\tau_2~.
\ee
The vector $A_\mu$ remains massless, and its dressed propagator will be approximated by its bare form (\ref{bareD}).
The propagator for the vector $B_\mu$ will be approximated by (\ref{DB}), where we consider the Feynman gauge $\zeta=0$, 
according to the discussion given in the introduction.\\
With these approximations, and taking into account the identities  $\tau_2\tau_3\tau_2=-\tau_3$ and $\tau_2^2=1$,
the Schwinger-Dyson (\ref{SD}) leads to
\bea
&&m_1+m_2\tau_3\\
&=&-im_1\int_p\frac{p^2-m_1^2+m_2^2}{4m_1^2p^2-(p^2+m_1^2-m_2^2)^2}\gamma^\mu(g_A^2D^A_{\mu\nu}+g_B^2D^B_{\mu\nu})\gamma^\nu\nn
&&+im_2\tau_3\int_p\frac{-p^2-m_1^2+m_2^2}{4m_1^2p^2-(p^2+m_1^2-m_2^2)^2}\gamma^\mu(g_A^2D^A_{\mu\nu}-g_B^2D^B_{\mu\nu})\gamma^\nu\nonumber~.
\eea
Identifying terms independent of $\tau_3$ and those proportional to $\tau_3$, and assuming that $m_1\ne 0$, $m_2\ne0$, 
one obtains the following gap equations after a Wick rotation
\bea\label{gap}
1&=&\int_q\frac{q^2+m_1^2-m_2^2}{4m_1^2q^2+(q^2-m_1^2+m_2^2)^2}
\frac{4g_A^2(q^2+m_B^2)+g_B^2\left( 4q^2+m_B^2\right)}{q^2(q^2+m_B^2) (1+(\vec q)^2/M^2)}\\
1&=&\int_q\frac{q^2-m_1^2+m_2^2}{4m_1^2q^2+(q^2-m_1^2+m_2^2)^2}
\frac{4g_A^2(q^2+m_B^2)-g_B^2\left( 4q^2+m_B^2\right)}{q^2(q^2+m_B^2) (1+(\vec q)^2/M^2)}~,\nonumber
\eea
where $q^2=p_0^2+(\vec p)^2$. In terms of the unknown quantities
\be
u=\frac{m_1}{m_2}~~~~\mbox{and}~~~~\mu=\frac{m_2}{M}~,
\ee
the gap equations to solve are then of the form 
\begin{equation}\label{gapeq}
16\pi^3=I_z(\mu,u),~\quad z=\pm1~, 
\end{equation}
where
\bea\label{Iz}
I_z(\mu,u)&=&\int_0^{2\pi}d\theta\sin^2\theta\int_0^\infty\frac{dt}{1+\mu^2 t\sin^2\theta}~
\frac{t+z(u^2-1)}{4u^2t+(t-u^2+1)^2}\nn
&&~~~~~~~~~\times\frac{4(g_A^2+zg_B^2)t+a^2(4g_A^2+zg_B^2)}{t+a^2}~.
\eea
In the previous  integral, $t=q^2/m_2^2$, and the paramerer $a^2$ is given by eq.(\ref{mB}):
\be\label{a2}
a^2=\frac{m_B^2}{m_2^2}=\frac{8}{3}\frac{g_B^2}{g_A^2-g_B^2}=\frac{8}{3}\frac{\alpha_B}{\alpha_A-\alpha_B}~,
~~~~~~\alpha_i\equiv g^2_i/4\pi~.
\ee
In what follows, we will assume that $\alpha_A>\alpha_B$, in order to be consistent with 
the calculation of the vector mass (\ref{mB}).

\subsection{Mass hierarchy}

We expect the dimensionless ratio $\mu$ to satisfy $\mu<<1$, and we therefore expand the integral equations 
(\ref{gapeq}) for small $\mu$. Also, since we are looking for a solution with mass hierarchy, we will investigate 
if it is possible to have $u\simeq1$, such that
the fermion masses (\ref{mamb}) satisfy $m_\chi<<m_\psi\simeq2m_2$. To see if this is possible, we set $u=1$ in the integral equations
(\ref{gapeq}), and we keep the dominant contribution only. We will then check the consistency of this regime, as far as the coupling constants 
are concerned. We have the two gap equations (\ref{gapeq}):
\be
16\pi^3\simeq\int_0^{2\pi}d\theta\sin^2\theta\int_0^{1/\mu^2}dt\frac{4\pi}{(t+4)(t+a^2)}
\left( 4(\alpha_A+z\alpha_B)t+a^2(4\alpha_A+z\alpha_B)\right)~,
\ee
from which we obtain ($z=\pm1$)
\be\label{agaga}
4\pi(4-a^2)=\left( 4(4-a^2)\alpha_A+z(16-a^2)\alpha_B\right) \ln\left(1+ \frac{1}{4\mu^2}\right) 
-3za^2\alpha_B\ln\left( 1+\frac{1}{a^2\mu^2}\right)~,
\ee
with solutions
\bea\label{solutions}
\ln\left( 1+\frac{1}{4\mu^2}\right)&\simeq&\frac{\pi}{\alpha_A}\\
\ln\left( 1+\frac{1}{a^2\mu^2}\right)&\simeq&\frac{\pi(16-a^2)}{3a^2\alpha_A}~.\nonumber
\eea
One can see from this result that the dimensionless dynamical mass $\mu$ depends only on the coupling $\alpha_A$
\be\label{mufinal}
\mu\simeq\exp\left(\frac{-\pi}{2\alpha_A}\right) ~.
\ee
The regime $u^2\simeq1$ is consistent provided the coupling constants satisfy the second equation (\ref{solutions}). 
One can easily notice that, in order of magnitude, $a^2 \simeq 4$ is a consistent approximation, since in that 
case the second of the equations (\ref{solutions}) reduces to the first.  On taking into account the expression 
(\ref{a2}), we observe that the condition $a^2 \simeq 4$  
translates to the following relation among the couplings 
\begin{equation}\label{condab}
3\alpha_A\simeq5\alpha_B~. 
\end{equation}
This is consistent with the condition $a^2<16$, required by the positivity of  the logarithm in the second equation 
(\ref{solutions}),
since this latter condition gives $6\alpha_A>7\alpha_B$. In practice, the above equation should only be understood as 
an order of magnitude approximation. We note at this point that any renormalisation group running of the couplings 
should respect this condition in the infrared regime, where dynamical mass generation takes place.

As a consequence, one can find a consistent regime $ u^2\simeq1$, for which
$m_\chi<<m_\psi \simeq 2 m_2$. From eq.(\ref{a2}), we also observe that the vector mass is $m_B\simeq2 m_\psi$. 

 At this point, we can give an estimate of the ratio $m_\chi/m_\psi$, by noting the following. For the calculation of the integrals
(\ref{Iz}), we have assumed that $u^2=1$, and concluded that the coupling constants must satisfy the relation (\ref{condab}). 
If we relax slightly this constraint, we can find a non-vanishing value for $\varrho \equiv u^2-1\simeq 4m_\chi/m_\psi$ by approximating
\bea\label{fineq}
I_+(\mu,u)-I_+(\mu,1)&\simeq&16\pi^2 \varrho \int_\varrho^\infty dt\frac{(4-t)[(\alpha_A+\alpha_B)t+4\alpha_B+\alpha_B]}{t(t+4)^3}\nn
I_-(\mu,u)-I_-(\mu,1)&\simeq&16\pi^2 \varrho \int_\varrho^\infty dt\frac{(4+3t)[(\alpha_A-\alpha_B)t+4\alpha_B-\alpha_B]}{t(t+4)^3}
\eea 
where we took into account that $a^2\simeq 4$. Since $I_+(\mu,u)=I_-(\mu,u)$, and we calculated $\mu$ above according to the 
identity $I_+(\mu,1)=I_-(\mu,1)$,
we conclude that the integrals appearing on the right-hand side of the equations (\ref{fineq}) must be equal. Evaluating 
these integrals analytically and identifying them leads to the following expression
\be
(4\alpha_A+\alpha_B)\left( \ln\left( \frac{1}{\varrho}\right) +2\ln2\right) -8\alpha_A+\alpha_B
=(4\alpha_A-\alpha_B)\left( \ln\left( \frac{1}{\varrho}\right) +2\ln2\right) +8\alpha_A-5\alpha_B~,
\ee
from which we finally obtain 
\be
\varrho = 4\exp\left( 6-8\frac{\alpha_A}{\alpha_B}\right) \simeq 2.6 \times 10^{-3} \ll 1~,
\ee
using the order of magnitude (\ref{condab}). 
Thus,  we consistently obtain the following  mass hierarchy:
\begin{equation}
m_\chi \simeq 1.3 \times 10^{-3} M e^{-\frac{\pi}{2\alpha_A}}  
 \ll m_\psi \simeq 2 M e^{-\frac{\pi}{2\alpha_A}}  < m_B \simeq  2 m_\psi = 4 M e^{-\frac{\pi}{2\alpha_A}}~.
\label{hierarchy} 
\end{equation}
We stress once more that this hierarchy should be considered together with 
the condition (\ref{condab}) between the (weak) couplings for consistency. 

The hierarchy we just found is thus in the right direction for more realistic models involving the entire Standard Model non-Abelian gauge group, 
which we hope to discuss in a forthcoming publication. It is important to stress, though, once again that the hierarchies derived 
here and 
in \cite{AM}, are relative hierarchies between the various masses. The absolute mass scale of the lightest excitations is very 
suppressed 
if we consider the LV models by themselves. It is only after embedding them to multiple brane models~\cite{branes}, with a 
reverse 
Randall-Sundrum hierarchy, that phenomenologically realistic values for the masses are obtained. At present we cannot avoid 
the use of  
such exotic enhancement mechanisms, and indeed this may be the only possibility for such LV models. In the subsection that 
follows, 
such embedding mechanism play an important r\^ole in understanding better the energetics of the mass-generated ground state of 
the models.

\subsection{Energetics}

The dynamical mass generation models of \cite{JackJohn},\cite{CornNort} are \emph{not} characterised by minimisation of the effective potential, 
unlike the Higgs-type models~\cite{higgs}. In fact, in \cite{CornNort} a generic argument was given that the energy of the massive ground 
state is the same to that of the massless one. The argument is based on the fact that in those field theories, which do not involve any mass 
scale initially, the generated fermion mass depends on a \emph{continuous }arbitrary mass parameter $M$. A detailed argumentation, using the 
quantum structure of the models 
has been provided in \cite{CornNort}, where we refer the interested reader for details. For our purposes here we invoke a simpler, but 
equally powerful proof based on a theorem independently proposed by several authors but known mostly by the name Feynman-Hellmann 
theorem~\cite{feynman}.
The theorem states that if there is a ground state $|\Psi_\lambda\rangle $ of a system with Hamiltonian that depends on a parameter $\lambda$, 
then for the energy $E$ of this ground state we have:
\begin{equation}\label{fhth}
\frac{\partial E}{\partial \lambda} = \langle \Psi_\lambda | \frac{\partial \widehat{H}}{\partial \lambda} | \Psi_\lambda \rangle~,
\end{equation} 
where $\widehat{H}$ is the Hamiltonian operator of the system. It is important that the parameter $\lambda$ is \emph{continuous}. 
The theorem uses, of course, in its original formulation the Schroedinger equation, 
but it suffices to demonstrate the main features of the models of interest to us here. Nevertheless, as already mentioned, the pertinent 
argumentation has been extended in \cite{CornNort} to incorporate field theoretic systems. In the dynamical models of \cite{CornNort}, 
the r\^ole of the continuous parameter is played by the mass scale $M$ apprearing in the dynamically generated masses, which is arbitrary. 
The Hamiltonian is independent of $M$, as there is no explicit mass scale in the Lagrangian, and hence, as a result of (\ref{fhth}), the 
ground state energy $E$ is independent of $M$. This implies that, in contrast to the Higgs case, where the mass generation implies a 
minimisation of the Higgs potential,  the ground state energy in the dynamical symmetry breaking Higgless models is independent of $M$ 
and hence it vanishes along with the massless case. 

Now we come to our models (\ref{bare}) or (\ref{model}). In contrast to the case of \cite{CornNort}, there is explicit dependence on the 
scale $M$ in the Hamiltonian of these models, due to the LV higher derivative terms. The result is 
\begin{equation}\label{fhthours}
\frac{\partial E}{\partial \lambda } = +\frac{1}{4}~ {_M }\langle 0 | \int d^4 x^E \left(F_{\mu\nu} \Delta F^{\mu\nu} 
+ G_{\mu\nu} \Delta G^{\mu\nu} \right)_E | 0 \rangle_M ~, \quad \lambda = M^{-2} ~.
\end{equation}
where the index $E$ denotes Euclidean formalism, as a result of the fact that the Hamiltonian  of the system is identified with minus the effective Euclidean action. 

One should expect that the Lorentz-violating nature of the vacuum $|0\rangle_M$ implies in general the non vanishing of the right-hand-side, 
implying a dependence of the vacuum energy on the dynamically generated mass.
Using the cyclic Bianchi identity for the gauge bosons field strengths, 
\begin{equation}\label{bianchi}
\partial_{\left[\mu\right.} F_{\left. \nu\rho\right]} = 0~, \quad \partial_{\left[\mu\right.} G_{\left. \nu\rho\right]} = 0~,
\end{equation}
with the symbol $[ \dots ]$ denoting symmetrisation of the appropriate indices, we obtain
\be
\frac{\partial E}{\partial \lambda } =- \frac{1}{4}~ {_M }\langle 0 | \int d^4 x^E \left(
F_{\mu\nu}\partial_i[\partial^\mu F^{\nu i}+\partial^\nu F^{i\mu}]+
G_{\mu\nu}\partial_i[\partial^\mu G^{\nu i}+\partial^\nu G^{i\mu}]\right)_E | 0 \rangle_M ~.
\ee
Integrating by part and assuming that the fields decay away at space-time infinity, one may write eq.(\ref{fhthours}) in the form:
\be\label{fhthours2}
\frac{\partial E}{\partial \lambda } = +\frac{1}{2} ~{_M} \langle 0 | \int d^4 x^E \left(\partial^\mu F_{\mu\nu} \partial_i F^{\nu i} 
+\partial^\mu G_{\mu\nu} \partial_i G^{\nu i} \right)_E | 0 \rangle_M
\ee
We write then the equations of motion for the vector fields, from the Lagrangian (\ref{model}) where we neglect the operator $\Delta/M^2$,
and we obtain:
\be\label{squarecharge}
\frac{\partial E}{\partial \lambda } 
= +\frac{1}{2} ~{ _M} \langle 0 | \int d^4 x \left((J_A^0)^2  + (J_B^0)^2
+ \vec J_A\cdot\vec J_A  + \vec J_B\cdot\vec J_B-J_{A,k}\partial_0F^{k0}-J_{B,k}\partial_0G^{k0} \right)_E|0\rangle_M~,
\ee
where the currents are $J^\mu_A = \overline{\psi} \gamma^\mu \psi$ and $J_B^{\mu} = \overline{\psi} \gamma^\mu \tau_2 \psi$ 
and the reader is reminded of the Euclidean formalism used in (\ref{squarecharge}).
The issue is whether the LV vacuum admits non trivial components of the vacuum currents for the fermions. If such 
fluctuations are absent in the theory, then the situation resembles that of \cite{CornNort}, in which the vacuum energy 
is independent of 
the scale $M$ and thus the vacuum energy for the case of non-trivial mass generation is degenerate with the massless case. 

From the point of view of embedding such LV field theories into a microscopic string/brane theory framework, as in \cite{branes}, 
such an energy degeneracy is consistent with the \emph{landscape} nature of string theory vacua. In fact, as we have discussed in 
some detail in \cite{branes}, to obtain the 
effective low-energy LV field theory of the generic type (\ref{bare}) a particular \emph{quantum ordering} of the higher derivative 
operators is required.
For one particular ordering the LV terms are absent. Various orderings correspond to different vacua, which conserve or violate 
Lorentz symmetry.
In the context of string landscape, such vacua are all degenerate in energy and it is only the requirement of a \emph{well-defined} 
low energy (\emph{i.e}. \emph{infra-red}) effective field theory \emph{limit} that seems to be the \emph{selection} criterion for 
the \emph{massive phase}, where infra-red infinities are absent.

Nevertheless, in the framework of the LV model studied here, one might face a situation where non-trivial condensates of  
squares of stationary four-currents $J^\mu_{A,B}$ are observed in the (rotationally invariant) vacuum. For such stationary 
currents, where $\partial_0 F^{k0}=\partial_0 G^{k0}=0$, we have then from eq.(\ref{squarecharge}):
\begin{eqnarray}
\frac{\partial E}{\partial \lambda } = \frac{1}{2}~_M\langle 0 | \int d^4 x \left( J_A^\mu J_{A\, \mu}  
+ J_B^\mu J_{B \, \mu} \right)_E |0\rangle_M \ge 0~.
\end{eqnarray} 
This implies that the vacuum energy $E$ in this case 
is a monotonically decreasing function of $M^2$, so that the vacuum energy of  
the massive phase  would be smaller than that of the massless case, and thus mass generation would be energetically preferable. 
Turning the logic around, we may also say that in the presence of higher-derivative Lorentz-violating terms in the action of the model, 
it is energetically preferable to have non-trivial vacuum condensates of the sum of squares of the currents in the massive phase of the model.

\section{Conclusions \label{sec:4}}

The model (\ref{model}) we considered here contains Lorentz-violating terms which are not measurable at the classical level, if 
the mass scale $M$ 
is of the order of the Plank mass. Nevertheless, quantum corrections generate finite effects from these Lorentz-violating terms, 
which lead to
the dynamical generation of masses for elementary particles (fermions and vectors). 
The mechanism of vector mass generation is based on the exchange of a massless
fermion-antifermion bound state, which represents a scalar excitation. This scalar decouples from the spectrum, 
due to a cancellation of the corresponding 
propagator pole with a singular effective vertex, resulting from a non-trivial fermion self energy structure, generated dynamically. 
Finally, the resulting masses exhibit a natural hierarchy.

However, as already emphasised in the text above, the \emph{toy models} discussed here, involving only vector 
interactions among the fermions, suffer from a serious phenomenological drawback: the deviation of the value of the maximal (light-cone) 
speed seen by the fermions (\ref{vpvg}) from that of the speed of light \emph{in vacuo} $c$, as a result of Lorentz violation, is unacceptably 
large to be compatible with the current phenomenology~\cite{coleman,sigl}, unless \emph{unnaturally small} couplings for the LV gauge groups are invoked. A possible way out would be to enhance the gauge group by including 
also axial-vector interactions~\cite{AM}, which, in view of their repulsive nature, may suppress the above deviation by terms proportional to 
the difference of the fine structure constants between the vector and axial-vector interactions. In general, such gauge groups are needed to 
describe the Standard Model Physics in this context, and one might be hopeful that the above described toy Lorentz-Violating models may have 
a chance of providing phenomenologically realistic theories, upon being extended to include axial vector interactions of Standard Model type. 

The extension of this work to a non-Abelian gauge theory needs to be looked at carefully, though. Indeed, in order not to break gauge 
invariance, 
higher-order space derivatives should naively be covariant, therefore introducing new interactions which are not renormalizable. 
To avoid this,
one needs to define a new field strength, where the higher-order space derivatives act on the Abelian sector only. A detailed 
analysis of such issues is postponed to a future publication, where more realistic phenomenological models will be studied.

Before closing we would like to make a final remark concerning the limit where the LV mass scale $M \to \infty$, which some readers might wish to take, especially if they attempt to view the current models as LV regulators~\footnote{ Moreover, as we have 
already mentioned, when we embed the model in more microscopic brane world models with higher-dimensional bulk spaces punctured 
by LV point-like brane structures, acting as defects, the scale $M$ is effectively inversely proportional to the density of 
these defects through their fluctuation $\sigma^2$ (\emph{cf.} (\ref{MMs})).
Hence in the limit of vanishing defects density we have an explicit realization of the limit $M \to \infty$.}.  
First of all, we should note that the above-mentioned finite maximal speed for fermions (\ref{vpvg}), which is \emph{independent} of the cut-off $M$, would persist in the $M \to \infty$ case and in this sense one cannot simply view the LV model as a mere regulator. There are physical effects that cannot removed in the limit $M \to \infty$. Moreover, in such a limit, 
the dynamically generated masses 
appear to diverge.  
However, one may envisage a Randall-Sundrum (RS) normal hierarchy situation in which a shadow world is placed 
at distance $\ell $ from our brane world. According to this picture, one assumes that on the shadow world the masses are generated 
by our LV Higgsless mechanism, and are all proportional to $M$. Due to the warped bulk space-time of the RS model, the corresponding 
masses in our world are all scaled by the factor~\cite{RS}:  $m \propto  M e^{-\kappa \ell} $, where $\kappa $ is proportional to 
the bulk gravitational constant of the higher dimensional bulk space time.
A limit $M \to \infty$ can then lead to finite results for the masses by letting $\ell \to \infty$ such that $M e^{-\kappa \ell} 
$= finite, of a value that sets a realistic mass scale on our world. This preserves the resulting hierarchy. Notice that it is only 
in the limiting case where the LV scale $M \to \infty$ that we apply the normal RS hierarchy. In this sense, the latter is viewed 
simply as an extra regularization. On the other hand, in the microscopic stringy models of \cite{branes}, with a \emph{finite} density of bulk space-time defects, where $M $ is kept \emph{finite}, one needs an inverted RS hierarchy to obtain realistic masses, as explained in that work.

\section{Acknowledgements} 

The authors wish to thank A. Pilaftsis and A. Vergou for discussions. 
The work of J.A. is supported in part by the Royal Society (UK), while that of N.E.M. is partially supported by the 
London Centre for Terauniverse Studies (LCTS), using funding from the European
Research Council via the Advanced Investigator Grant 267352.

\end{document}